# Identification of an ultrafast internal conversion pathway of pyrazine by time-resolved vacuum ultraviolet photoelectron spectrum simulations

Manabu Kanno,[1,a] Benoît Mignolet,[2] Françoise Remacle,[2] and Hirohiko Kono[1]

[1] *Department of Chemistry, Graduate School of Science, Tohoku University, Sendai 980-8578, Japan*

[2] *Theoretical Physical Chemistry, UR MOLSYS, University of Liège, B4000 Liège, Belgium*

The internal conversion from the optically bright $S_2$ ($^1B_{2u}$, $\pi\pi^*$) state to the dark $S_1$ ($^1B_{3u}$, $n\pi^*$) state in pyrazine is a standard benchmark for experimental and theoretical studies on ultrafast radiationless decay. Since 2008 a few theoretical groups have suggested significant contributions of other dark states $S_3$ ($^1A_u$, $n\pi^*$) and $S_4$ ($^1B_{2g}$, $n\pi^*$) to the decay of $S_2$. We have previously reported the results of nuclear wave packet simulations [Phys. Chem. Chem. Phys. **17**, 2012 (2015)] and photoelectron spectrum calculations [Chem. Phys. **515**, 704 (2018)] that support the conventional two-state picture. In this article, the two different approaches, i.e., wave packet simulation and photoelectron spectrum calculation are combined: We computed the time-resolved vacuum ultraviolet photoelectron spectrum and photoelectron angular distribution for the ionization of the wave packet transferred from $S_2$ to $S_1$. The present results reproduce almost all the characteristic features of the corresponding experimental time-resolved spectrum [T. Horio *et al.*, J. Chem. Phys. **145**, 044306 (2016)] such as a rapid change from a three-band to two-band structure. This further supports the existence and character of the widely accepted pathway ($S_2 \rightarrow S_1$) of ultrafast internal conversion in pyrazine.

[a] Author to whom correspondence should be addressed: manabu.kanno.d2@tohoku.ac.jp

## I. INTRODUCTION

A variety of heteroaromatic molecules (e.g., DNA and RNA nucleobases) is known to undergo ultrafast radiationless decay after ultraviolet (UV) excitation.[1–6] The mechanism of ultrafast radiationless decay has been intensively investigated in simpler prototypical systems such as pyrazine for decades.[7–9] Evidence of the ultrafast radiationless decay in pyrazine was identified in diffuse bands in its UV absorption spectrum.[10] Domcke and coworkers theoretically explored its potential energy surfaces within a restricted-vibrational-mode model and proposed that the ultrafast radiationless decay is due to the internal conversion (IC) from the optically bright $S_2$ ($^1B_{2u}$, $\pi\pi^*$) state to the almost dark $S_1$ ($^1B_{3u}$, $n\pi^*$) state via a conical intersection (CI).[11–13] This two-state picture explains many experimental results reported so far and thus has been widely accepted. Within this picture the $S_2$ lifetime was estimated to be about 20 fs by femtosecond time-resolved photoelectron spectroscopy.[14,15] The IC between the two states in pyrazine has become a standard benchmark for testing the performance of theoretical approaches to describe or control nonadiabatic dynamics in polyatomic molecules.[16–23]

Since 2008, a few theoretical groups have questioned the well-established IC pathway of pyrazine.[24–30] Werner *et al*. computed temporal change in the populations not only of $S_1$ and $S_2$ but also of two other optically dark $n\pi^*$ states, $S_3$ ($^1A_u$, $n\pi^*$) and $S_4$ ($^1B_{2g}$, $n\pi^*$), by combining the time-dependent density functional theory (TDDFT) with Tully's semiclassical fewest switches surface hopping procedure.[24] In fact, the two dark states $S_3$ and $S_4$ were observed by spectroscopic methods other than UV absorption.[31,32] We designate $S_1$ to $S_4$ in the order of their measured excitation energies. Surprisingly, Werner *et al*. found that $S_3$ and $S_4$ acquire almost 60 % of the initial $S_2$ population in total soon after optical excitation (~ 10 fs) followed by a gradual rise of $S_1$. They have proposed a four-state ($S_1$ to $S_4$) picture of ultrafast IC in pyrazine. Sala *et al*. also included the four excited states in their multi-configuration time-dependent Hartree (MCTDH) calculation with a 16-mode model Hamiltonian based on the extended multi-configuration quasi-degenerate second-order perturbation theory (XMCQDPT2).[27] In contradiction to both the two- and four-state pictures, their results suggested a three-state ($S_1$ to $S_3$) picture in which a nonadiabatic transition from $S_2$ to $S_3$ (but not to $S_4$) competes with that from $S_2$ to $S_1$. If either of the four- or three-state pictures were true, previous findings accumulated by countless experimental and theoretical studies focused only on $S_1$ and $S_2$ might be subject to re-examination.

With reference to spectroscopic data for vertical excitation energies[31,32] we pointed out that both TDDFT and XMCQDPT2 severely underestimate the energy of $S_3$ that can be even computed below $S_2$ and that the former method also locates $S_4$ too close to $S_2$.[33] As a consequence, in the two methods, pyrazine encounters the CI with $S_3$ (and that with $S_4$ in the TDDFT case) in the course of propagation in $S_2$. Furthermore, we carried out quantum nuclear wave packet (WP) dynamics simulations in a minimal two-vibrational-mode model using the highly accurate multi-reference configuration interaction (MRCI) method.[33] Our simulations demonstrated that nonadiabatic transitions to $S_3$ and $S_4$ are negligible and thereby supported the conventional direct two-state ($S_2 \rightarrow S_1$) picture. However, Sala *et al.* claimed that the reference value for the vertical excitation energy to $S_3$ is controversial and that a large amount of the population is transferred to $S_3$ even when its energy is shifted above $S_2$ in their higher-dimensional calculation.[28]

In order to conclusively clarify an ultrafast IC pathway of photo-excited pyrazine, it seems necessary to compare simulation results with latest reliable experimental data. We especially pay attention to the time-resolved photoelectron imaging experiment with a 9.3-eV vacuum UV (VUV) probe pulse reported by Horio *et al*.[34] Photoelectron spectroscopy enables the detection of dark states. Rich dynamical information of molecules can be imprinted in the observed image by ionizations up to high cationic states with a VUV photon. The measured time-resolved photoelectron spectrum (PES) of pyrazine exhibited a dramatic change from a three-band to two-band structure arising from ultrafast IC as a function of the pump-probe delay time. They also extracted the anisotropy parameter[35,36] from the observed photoelectron angular distribution (PAD). No signature of $S_3$ or $S_4$ was detected in the experiment.

In our recent study, as a first step of a theoretical analysis of the pump-probe photoelectron imaging experiment by Horio *et al*., we computed VUV PESs for ionizations from each of the four neutral excited states ($S_1$ to $S_4$) under the sudden-ionization approximation.[37] The PESs were obtained by a single-point calculation: We evaluated that from $S_2$ at the Franck-Condon (FC) position and those from the other three excited states at the respective potential minima. The resultant PESs from $S_2$ and $S_1$ are analogous to those observed at short and long delay times, respectively, whereas those from $S_3$ and $S_4$ exhibit a peak at totally different positions from the experimental ones.

In the present study, to further strengthen our previous findings, we take the next step beyond a single-point calculation and also beyond the sudden-ionization approximation: The time-resolved VUV PES and PAD (anisotropy parameter) are computed taking into account the propagation of nuclear WPs reported in Ref. 33 and the durations of pump and probe pulses. This realizes a more detailed analysis of the rapidly varying multi-band structure of the measured PES. From the analysis, we identify ionization channels contributing to individual spectral bands and add in our modeling a couple of channels that were not considered either in the experimental assignments[34] or in our earlier study.[37] Using this extended model for ionization channels, almost all the characteristic features of the measured PES can be interpreted as a result of IC from $S_2$ to $S_1$, which provides more solid support for the conventional two-state picture of ultrafast IC in pyrazine.

The remainder of this article is organized as follows. In Sec. II, we describe the methods for our electronic structure computation, quantum nuclear WP simulation, and time-resolved

PES calculation. In Sec. III, the resultant PES and anisotropy parameter are compared to the respective experimental ones. Finally, Section IV concludes this article.

## II. COMPUTATIONAL METHOD

### A. Electronic structure and nuclear WPs of neutral pyrazine

For neutral pyrazine, we utilize the results of our previous electronic structure calculations and quantum nuclear WP simulation reported in Ref. 33. A summary of the computational methods used is provided below. The $D_{2h}$ optimized geometry and harmonic vibrational normal modes of pyrazine in the electronic ground state $S_0$ ($1^1A_g$) were obtained at the state-averaged complete-active-space self-consistent field (SA-CASSCF)[38,39] level of theory with an active space of ten electrons in eight orbitals (three $\pi$, three $\pi^*$, and two lone-pair orbitals). The SA-CASSCF energies were then refined at the level of the internally contracted MRCI including single and double excitations to the external space (MRCISD).[40–42] These calculations were performed with the 6-311++G** Gaussian basis set[43] by using the *ab initio* quantum chemistry software MOLPRO.[44,45] The SA-CASSCF/MRCISD treatment overestimated the measured vertical excitation energies to $S_1$ ($1^1B_{3u}$), $S_2$ ($1^1B_{2u}$), $S_3$ ($1^1A_u$), and $S_4$ ($1^1B_{2g}$) but reproduced their relative energies much better than TDDFT and XMCQDPT2, which gave inconsistent results with spectroscopic data (see Table 3 in Ref. 33). Needless to say, an electronic structure method that can accurately replicate all the measured excitation energies is ideal for conclusive analysis; however, to the best of our knowledge, there has so far been no such report. As a reasonable compromise, the MRCISD energies of the neutral excited states were shifted (lowered) by 0.53 eV as in Ref. 37.

To construct a minimum-vibrational-mode model for the early stage of IC in pyrazine, out of its 24 normal modes we chose two often called $Q_{6a}$ and $Q_{10a}$. The former is a totally symmetric ($a_g$) in-plane ring deformation mode to which a vibrational quantum beat observed in photoelectron signals of pyrazine was ascribed;[15] the latter involves an out-of-plane CH bending vibration and is the only $b_{1g}$ mode that couples $S_1$ ($1^1B_{3u}$) and $S_2$ ($1^1B_{2u}$) (see Fig. 1 in Ref. 33 for the vibrational vectors of $Q_{6a}$ and $Q_{10a}$). In the two-dimensional (2D) space spanned by $Q_{6a}$ and $Q_{10a}$, the symmetry of pyrazine is lowered to $C_{2h}$ and the two excited states belong to $^1B_u$. Along the normal coordinates obtained by SA-CASSCF, $\mathbf{Q} \equiv (Q_{6a}, Q_{10a})$, we uniformly distributed 4096 (64 × 64) grid points in the square region with a side of 10 $u^{1/2}a_0$, i.e., [−5 $u^{1/2}a_0$, 5 $u^{1/2}a_0$] for both $Q_{6a}$ and $Q_{10a}$, where u is the unified atomic mass unit and $a_0$ is the Bohr radius. From symmetry consideration, the MRCISD calculations were executed only in the region with $Q_{10a} \geq 0$. The computational cost was further reduced: The energies of $S_1$ ($1^1B_u$) and $S_2$ ($2^1B_u$) were evaluated at every other point along $Q_{6a}$ and $Q_{10a}$, followed by cubic spline interpolation for the intermediate points.

For exploration of IC dynamics in the 2D model, the vibrational ground-state wave function in $S_0$ was vertically excited to $S_2$ at the initial time $t = 0$. The subsequent evolution of the FC WP was simulated in the diabatic basis to deal with nonadiabatic couplings using the quasi-diabatization scheme implemented in MOLPRO.[46] The diabatic WPs were propagated by numerically integrating their coupled differential equations and then transformed into adiabatic WPs $\chi_J(\mathbf{Q}, t)$ with $J$ designating $S_1$ or $S_2$. Finally, the probability densities $|\chi_J(\mathbf{Q}, t)|^2$ were obtained at the 4096 grid points. The resultant decay curve of the $S_2$ population was analogous to that in the calculation with a 24-mode model Hamiltonian by Puzari et al.[18] until $t \sim 80$ fs, which corroborates the predominance of $Q_{6a}$ and $Q_{10a}$ over the remaining 22 modes in this time range (see Fig. 10 in Ref. 33).

**B. Electronic structure of cationic pyrazine**

Photoionization cross sections at all the 4096 grid points are not necessary to achieve a sufficiently accurate PES for comparison with the experimental one. The smaller square region with a side of 5.625 $u^{1/2}a_0$ is enough to conserve the total population in $S_1$ and $S_2$. We divide the region into 81 (9 × 9) unit cells and take account of cross sections at their central positions (i.e., 0, ±0.625, ±1.25, ±1.875, and ±2.5 $u^{1/2}a_0$ for both $Q_{6a}$ and $Q_{10a}$), which are part of the 4096 points. The cross sections at the 81 cell centers are weighted by the WP probability densities integrated over individual cells as will be expressed later in Eq. (6). In fact, one only needs cationic states at 45 (9 × 5) points with $Q_{10a} \geq 0$ from symmetry consideration.

At the 45 points we evaluated two $^2A_g$, two $^2B_g$, and five $^2B_u$ states relevant to the ionizations of $S_1$ and $S_2$ with a 9.3-eV photon of the VUV probe pulse employed in the experiment by Horio et al.[34] The same SA-CASSCF/MRCISD treatment as for neutral states was adopted for each symmetry species except that the number of active electrons was decreased to nine. The MRCISD energies of the cationic states are in excellent agreement with the measured ionization potentials (IPs) from $S_0$ (see Table 1 in Ref. 37). See also Table 3 in Ref. 37 for the order and dominant electronic configurations of the states at the FC position **Q** = 0.

## C. PES simulation

### 1. Photoionization matrix elements and PAD

Photoionization matrix elements[47,48] are a key quantity in the computation of photoionization cross sections, PADs, and PESs. They are expressed as the dipolar coupling between the wave function of a neutral electronic state, $\Psi_J^{neut}(\mathbf{Q})$, and that of an ionized state:[49]

$$\mathbf{D}_{IK\varepsilon\Omega}(\mathbf{Q}) = \langle \Psi_J^{neut}(\mathbf{Q}) | \boldsymbol{\mu} | \Psi_K^{cat}(\mathbf{Q}), \Psi_{\varepsilon\Omega}^{elec} \rangle, \tag{1}$$

where **μ** is the electric dipole moment operator. The ionized state is defined as the antisymmetrized product of the wave function of a cationic state, $\Psi_K^{cat}(\mathbf{Q})$, and that of the ionized electron, $\Psi_{\varepsilon\Omega}^{elec}$, where $\varepsilon$ and $\Omega$ denote the photoelectron kinetic energy (PKE) and emission solid angle, respectively. We chose to describe $\Psi_{\varepsilon\Omega}^{elec}$ by a plane wave.[48–51] The $n$-electron integral of Eq. (1) can be reduced to a one-electron integral using Dyson orbitals, $\phi_{JK}^{Dyson}(\mathbf{Q};\mathbf{r})$, i.e., the overlap between the neutral and cationic electronic states:[47,48,52]

$$\mathbf{D}_{JK\varepsilon\Omega}(\mathbf{Q}) = \langle \phi_{JK}^{Dyson}(\mathbf{Q};\mathbf{r}) | \boldsymbol{\mu} | \Psi_{\varepsilon\Omega}^{elec}(\mathbf{r}) \rangle, \qquad (2)$$

where $\mathbf{r}$ is the electronic coordinate. Its explicit expression is

$$\phi_{JK}^{Dyson}(\mathbf{Q};\mathbf{r}) = \sqrt{n} \int \Psi_J^{neut}(\mathbf{Q};\mathbf{r},\mathbf{r}_2,...,\mathbf{r}_n) \Psi_K^{cat}(\mathbf{Q};\mathbf{r}_2,...,\mathbf{r}_n) d\mathbf{r}_2...d\mathbf{r}_n. \qquad (3)$$

The angle-resolved photoionization cross section is given by[48,49]

$$\frac{d\sigma_{JK}(\mathbf{Q},\varepsilon,\Omega)}{d\Omega} = \frac{8\pi\omega}{\varepsilon_0 c} \rho_\Omega(\varepsilon) |\mathbf{e} \cdot \mathbf{D}_{JK\varepsilon\Omega}(\mathbf{Q})|^2, \qquad (4)$$

where $\varepsilon_0$ is the vacuum permittivity, $c$ is the speed of light, $\omega$ is the probe carrier frequency, $\mathbf{e}$ is the electric field polarization vector, and $\rho_\Omega(\varepsilon)$ is the density of plane waves. The PAD is proportional to the angle-resolved cross section and the total cross section $\sigma_{JK}(\mathbf{Q}, \varepsilon)$ can be obtained by integrating it over $\Omega$. The value of $\varepsilon$ was sampled with a step of 0.05 eV and 600 emission angles were used for each PKE. Since molecules are randomly oriented in the experiment of interest, an averaging over molecular orientations is needed.[53] If an electric field was applied along a fixed direction to a set of randomly oriented molecules, the calculation of photoionization matrix elements for each orientation of the molecule would be very time-consuming. Instead, we applied a randomly oriented electric field to a fixed molecule; once $\mathbf{D}_{JK\varepsilon\Omega}(\mathbf{Q})$ is computed for the molecule, taking its inner product with $\mathbf{e}$ is easy. The resultant PAD was then rotated using Euler angles[54] so that the electric field is polarized along the $z$ axis and the molecule is randomly oriented. Overall 6000 random orientations were sampled.

The PAD of randomly oriented molecules along the polar angle $\theta$ between the directions of laser polarization ($z$ axis) and photoelectron emission (wave vector) is reflected by the anisotropy parameter $\beta$,[35,36] which varies between +2 and −1. A positive (negative) value of $\beta$ indicates that photoelectrons are mainly ionized parallel (perpendicular) to the polarization direction; $\beta = 0$ for an isotropic PAD. We first integrated a PAD over the azimuthal angle to obtain the intensity as a function of $\theta$, $I_{JK}(\mathbf{Q}, \varepsilon, \theta)$, and then fitted it with the equation below:

$$I_{JK}(\mathbf{Q},\varepsilon,\theta) = \frac{\sigma_{JK}(\mathbf{Q},\varepsilon)}{4\pi}\left[1 + \beta_{JK}(\mathbf{Q},\varepsilon)P_2(\cos\theta)\right], \tag{5}$$

where $P_2(\cos\theta)$ is the second-order Legendre polynomial.

### 2. Time-resolved PES

We first computed the PES corresponding to the sudden ionization[52] of the $S_1$ and $S_2$ WPs at each $t$. The sudden-ionization approximation assumes the occurrence of ionizations at the maximum of the pulse envelope. Under this approximation, the PES depends on the IPs and total photoionization cross sections at the sampled grid points and on the probability densities of the neutral WPs:

$$P^{Stick}(t,\varepsilon) \propto \sum_{i=1}^{81}\sum_{J}\sum_{K} \delta\left(\hbar\omega - IP_{JK}(\mathbf{Q}_i) - \varepsilon\right)\sigma_{JK}(\mathbf{Q}_i,\varepsilon)C_{Ji}(t), \tag{6}$$

where $\delta$ is the Dirac delta function, $\hbar$ is the Dirac constant, $IP_{JK}$ is the vertical IP, $\mathbf{Q}_i$ is the center of the $i$th unit cell in the 2D space, and $C_{Ji}(t)$ is the time-dependent probability density $|\chi_J(\mathbf{Q}, t)|^2$ integrated over the $i$th cell. The PES in Eq. (6) is constituted by a set of discrete peaks (stick PES) at each $t$.

In the results presented below, the stick PES, $P^{Stick}(t, \varepsilon)$, was convoluted by two Gaussian functions: One in energy and the other in time.[55] We define $P^{SI}(t, \varepsilon)$ as the PES obtained by the convolution with a Gaussian energy window function:

$$P^{SI}(t,\varepsilon) = \int P^{Stick}(t,\varepsilon') \frac{\exp\left[-(\varepsilon-\varepsilon')^2/2s^2\right]}{s\sqrt{2\pi}} d\varepsilon', \qquad (7)$$

where we set $s = 0.27$ eV equivalent to the full width at half maximum (FWHM) of 0.64 eV. The energy broadening allows for the bandwidth of the probe pulse. In this article the PES in Eq. (7) is referred to as the sudden-ionization PES.

Next, the sudden-ionization PES, $P^{SI}(t, \varepsilon)$, was also convoluted by a Gaussian time window function with a FWHM of 25.8 fs (standard deviation $S = 11$ fs):

$$P(t,\varepsilon) = \int P^{SI}(t',\varepsilon) \frac{\exp\left[-(t-t')^2/2S^2\right]}{S\sqrt{2\pi}} dt'. \qquad (8)$$

The time broadening considers the experimental cross correlation between pump and probe pulses (with a FWHM of 25 fs).[34] The time-resolved PAD (anisotropy parameter) was evaluated in the same manner as the time-resolved PES, $P(t, \varepsilon)$, in Eq. (8).

## III. RESULTS AND DISCUSSION

### A. IPs as a function of the $Q_{6a}$ and $Q_{10a}$ modes

Out of the three factors that contribute to the PES formula in Eq. (6), we first show the dependence of vertical IPs on the 2D nuclear coordinates **Q**; the other two factors, namely, cross sections (i.e., Dyson orbitals and PADs) and probability densities of neutral WPs will be presented in Sec. IIIB along with the computed PES. As seen in Eq. (6), the probe pulse can ionize the neutral WPs only to cationic states with an IP lower than its photon energy (9.3 eV in this study). The IPs for the ionizations from $S_1$ and $S_2$ to the nine cationic states included in our simulations are plotted in Fig. 1 as a function of the $Q_{6a}$ and $Q_{10a}$ modes. A color code is used to highlight which cationic states can be energetically accessed with a 9.3-eV photon; green (red) is for an IP sufficiently lower (higher) than 9.3 eV. In orange are ionization channels with an IP slightly higher than 9.3 eV, which can still be exceeded with the bandwidth of the probe pulse (0.21 eV) as expressed in Eq. (7).

In the following, we denote the nine cationic states by $1^2A_g$, $2^2A_g$, and so on but the two neutral states by $S_1$ and $S_2$ rather than by $1^1B_u$ and $2^1B_u$. One can clearly see in Fig. 1 that the lowest two $^2A_g$ states are reachable from both $S_1$ and $S_2$ irrespective of where the neutral WPs are localized in the 2D space. For $^2B_g$ and $^2B_u$ states, we should emphasize the strong variations of IPs for the $2^2B_g \leftarrow S_2/S_1$ and $1^2B_u \leftarrow S_2/S_1$ channels: The IPs drastically decrease with increasing $Q_{6a}$. In particular, the $2^2B_g \leftarrow S_2/S_1$ ionizations turn from closed to open channels as $Q_{6a}$ increases. These features of IPs should be reflected in the computed PES, which will be discussed in Sec. IIIB.

## B. Nonadiabatic dynamics and sudden-ionization PES

Figure 2 shows the time evolution of neutral WPs [to be precise, $C_{Ji}(t)$ in Eq. (6)] reported in Ref. 33 and the PES resulting from the sudden ionization of the WPs, $P^{SI}(t, \varepsilon)$, defined in Eq. (7). The sudden-ionization PES allows us to analyze all the features of PES in details, some of which will be washed out once the durations of pump and probe pulses are considered as in Eq. (8). Before carefully looking at the change in the sudden-ionization PES, let us briefly review nonadiabatic dynamics of the WPs in Fig. 2. The FC WP initially centered at $\mathbf{Q} = 0$ in $S_2$ moves in the negative direction of $Q_{6a}$ along the potential gradient of $S_2$. It is then transferred to $S_1$ around the CI located at $\mathbf{Q} = (-1.21\ u^{1/2}a_0, 0)$ by a nonadiabatic transition. The WP of $S_1$ propagates in the positive direction of $Q_{6a}$ along the $S_1$ gradient while bifurcating equally into two portions delocalized in the positive and negative regions of $Q_{10a}$. With reference to these WP motions we assign each peak in the sudden-ionization PES at selected time points in Fig. 2 ($t = 0, 13, 25, 37$, and $49$ fs).

At the initial time $t = 0$, the WP excited to $S_2$ is localized in the FC region and the PES is composed of three major bands. The band at $\varepsilon \sim 0.3$ eV stems from a combination of the $3^2B_u \leftarrow S_2$ and $2^2B_u \leftarrow S_2$ channels, and those at $\varepsilon = 2.5$ and 4.1 eV are ascribed to $1^2B_g \leftarrow S_2$ and $2^2A_g \leftarrow S_2$, respectively. The positions and assignments of the bands agree well with those for the PES obtained by a single-point calculation at $\mathbf{Q} = 0$ in Ref. 37. In addition, there is a weak side peak at $\varepsilon = 1.1$ eV, which was absent in the PES in Ref. 37. It originates from the $1^2B_u \leftarrow S_2$ channel. At $\mathbf{Q} = 0$, this ionization is forbidden because the norm of its Dyson orbital is almost zero and the photoionization matrix elements are thus very small (see Table 3 in Ref. 37 for electronic configurations of the states). At a neighboring position $\mathbf{Q} = (0, 0.625\ u^{1/2}a_0)$ (the center of an adjacent cell), the probability density of $S_2$ is however not zero and the Dyson orbital for $1^2B_u \leftarrow S_2$ has a norm of 0.16, which explains the presence of the weak side peak in the PES. The Dyson orbitals and PADs for the five ionization channels are displayed in Fig. 3. The PADs for $2^2B_u \leftarrow S_2$ and $1^2B_u \leftarrow S_2$ exhibit an isotropic distribution of photoelectrons with $\beta = 0.10$ and $-0.02$, respectively. In contrast, those for the other three channels are somewhat anisotropic; the $3^2B_u \leftarrow S_2$ channel produces a higher ionization yield parallel to the laser polarization direction along the $z$ axis ($\beta = 0.24$), whereas the $1^2B_g \leftarrow S_2$ and $2^2A_g \leftarrow S_2$ channels emit more photoelectrons perpendicular to the $z$ axis ($\beta = -0.29$ and $-0.35$, respectively).

At $t = 13$ fs, the WP of $S_2$ runs in the negative direction of $Q_{6a}$ and approaches the CI region. This WP motion induces an increase of the IPs for the ionizations from $S_2$ to $1^2B_u$, $2^2B_u$, and $3^2B_u$. The PES still consists of three bands but the first one (at the lowest $\varepsilon$) is now attributed only to $1^2B_u \leftarrow S_2$, of which $\varepsilon$ is red-shifted from 1.1 eV.

At $t = 25$ fs, a substantial part of the $S_2$ WP is transferred to $S_1$ by a nonadiabatic transition and the $S_1$ WP spreads in a wide range of $Q_{10a}$. Consequently, the first and second bands have almost disappeared and the PES is mainly composed of a single band arising from the $2^2A_g \leftarrow$ $S_2$ and $1^2A_g \leftarrow S_1$ channels. The values of $\varepsilon$ for these ionization channels are similar to each other as the energy gap between $1^2A_g$ and $2^2A_g$ is close to that between $S_1$ and $S_2$. Note that the intensity of the third band is also reduced. The norms of the Dyson orbitals for $2^2A_g \leftarrow S_2$ and $1^2A_g \leftarrow S_1$ are plotted in Fig. 4. One can see that they are both relatively small in the region where the $S_1$ WP is localized ($Q_{6a} < 0$ excluding $Q_{10a} \sim 0$). This accounts for the attenuation of the third band.

At $t = 37$ fs, the WP of $S_1$ evolves in the positive direction of $Q_{6a}$. This leads to a recovery of the third band since the Dyson orbital norm for $1^2A_g \leftarrow S_1$ rises with increasing $Q_{6a}$ as seen in Fig. 4. Moreover, a peak emerges at low $\varepsilon$ owing to a decrease of the IP for $1^2B_u \leftarrow S_1$.

At $t = 49$ fs, the $S_1$ WP keeps moving in the same direction as that at $t = 37$ fs and hence the IP for $1^2B_u \leftarrow S_1$ continues to drop as well as that for $2^2B_g \leftarrow S_1$. This brings about a blue shift of the peak for $1^2B_u \leftarrow S_1$ (to $\varepsilon = 1.1$ eV) and gives birth to a near-threshold peak for $2^2B_g \leftarrow S_1$. The latter peak was not obtained by a single-point calculation at the potential minimum of $S_1$, $\mathbf{Q} = (0.48\ u^{1/2}a_0, 0)$, in Ref. 37. The $2^2B_g$ state is accessible from $S_1$ only at $Q_{6a} > 1.5$ $u^{1/2}a_0$ (see Fig. 1) where a considerable part of the $S_1$ WP is localized (see Fig. 2). The Dyson orbitals and PADs for the three ionization channels are depicted in Fig. 5. The value of $\beta$ is negative for all the three PADs and they are more anisotropic (along the $z$ axis) with higher $\varepsilon$.

The PES formulas presented in Sec. IIC do not involve interference effects between ionization channels, which may in principle appear in the vicinity of the CI. In the CI region, the ionization to $3^2B_u$ is allowed from both $S_1$ and $S_2$ according to their electronic configurations (see Table 3 in Ref. 37) but as mentioned above the two ionization channels are energetically closed. No significant interference is thus expected and we have indeed confirmed numerically that the contribution of interference terms to the PES is negligible (less than 0.1 %).

## C. Time-resolved PES

For comparison with the experimental results reported by Horio *et al.*,[34] we evaluated the time-resolved PES, $P(t, \varepsilon)$, taking into account the durations of pump and probe pulses as in Eq. (8). Figures 6(a) and 6(b) display the measured and calculated time-resolved PESs, respectively. The computed PES consists of three bands as the experimental one. In both Figs. 6(a) and 6(b), the first and third bands grow with the pump-probe delay time *t*, while the second one quickly fades out. Unlike in the sudden-ionization PES in Fig. 2, the first and third bands in the time-resolved PES in Fig. 6(b) do not attenuate at $t < 25$ fs. This suggests that the experimentally observed continuous rise of the two bands comes from the time broadening with the cross correlation between pump and probe pulses. The decay of the second band is caused by a nonadiabatic transition through the CI and the $S_2$ lifetime can therefore be estimated from it; in Ref. 34 Horio *et al.* actually confirmed the agreement between the decay time (25 fs) and the $S_2$ lifetime (22 fs) determined in their previous experiment with a UV probe pulse.[15] As described in Sec. IIIB, the first band corresponds to the ionizations from $S_2$ to $1^2B_u$, $2^2B_u$, and $3^2B_u$ at $t < 25$ fs and to those from $S_1$ to $2^2B_g$ and $1^2B_u$ at $t > 25$ fs, resulting in a broadening of the band (mainly due to the blue shift in $1^2B_u \leftarrow S_1$). A similar trend is also observed experimentally in Fig. 6(a).

Overall the calculated time-resolved PES well reproduces the measured one except for two minor issues. One is the underestimation of the first band due to the plane-wave approximation for a photoelectron wave function. This approximation ignores the interaction between the outgoing electron and the cation core, which is particularly important at low $\varepsilon$. The observed intensity ratio between the three bands may be retrieved by multiplying the PES by an appropriate decreasing function $f(\varepsilon)$, which is yet unknown. Here, just for reference, let us assume a simple exponential function $f(\varepsilon) = \exp(-\alpha\varepsilon)$ to phenomenologically mimic the observed intensity ratio. In Fig. 6(c), we show the PES multiplied by $\exp(-\alpha\varepsilon)$ with $\alpha = 0.5$

eV$^{-1}$, in which the intensity ratio is also consistent with the experimental one in Fig. 6(a). The other issue is that a red shift of the third band is absent in the computed PES in Fig. 6(b). This has already been explained in Ref. 37: Shifting IPs for all ionization channels by 0.53 eV properly corrects that for $2^2A_g \leftarrow S_2$ but still underestimates that for $1^2A_g \leftarrow S_1$ by about 0.2 eV.

The temporal behavior of the three bands is clearly depicted in a time-energy 2D map of the PES in Fig. 7. Except for the underestimation of the first band, the calculated 2D map in Fig. 7(b) is in good accord with the measured one in Fig. 7(a). If we scale the computed PES by an exponential function, the agreement is even better [see Fig. 7(c)]. The results in Figs. 6 and 7 validate our quantum nuclear WP simulation in the conventional two-state ($S_2 \rightarrow S_1$) picture.

### D. Time-resolved anisotropy parameter

In addition to the time-resolved PES, we also calculated the variation of the anisotropy parameter $\beta$ including the effects of pump and probe pulse durations. Figures 8(a) and 8(b) show the experimental and computed time-energy 2D maps of $\beta$, respectively. One can confirm in Fig. 8(b) that the value of $\beta$ for each ionization channel in Figs. 3 and 5 is reflected in the map at $t \sim 0$ and 50 fs, respectively. Unfortunately it is difficult to conduct a thorough quantitative comparison with the experimental one for a couple of reasons. First, the error in the measured $\beta$ is estimated to be about ±0.1,[56] which is rather large compared to the maximum absolute value of $\beta$ (about 0.4) in Fig. 8. Second, Horio *et al.* detected photoelectron signals from Rydberg states produced by a nonresonant excitation with a 9.3-eV photon at negative and near-zero $t$.[34,56,57] The Rydberg signals are weak but tend to have large anisotropy (large positive $\beta$). Therefore the positive values of $\beta$ observed at $t \sim 0$ in Fig. 8(a) are likely to originate from the Rydberg signals [see Fig. 3(e) in Ref. 34]. Third, between the three bands their long tails overlap each other in the experimental PES in Figs. 6(a) and 7(a), while the bands are nearly separated in the calculated PES in Figs. 6(b) and 7(b). For these reasons we make a qualitative comparison between Figs. 8(a) and 8(b) focused on the specific region of $\varepsilon$ in which each band takes the maximum intensity.

For the first band, the measured $\beta$ in Fig. 8(a) is relatively large at very low $\varepsilon$ (< 0.2 eV), while it is small at 0.2 eV < $\varepsilon$ < 1 eV. The computed $\beta$ in Fig. 8(b) is also close to zero (−0.15 to +0.15) at $\varepsilon$ < 1 eV for all $t$. Given the error of about ±0.1 in the measured $\beta$, a reasonable agreement is achieved at 0.2 eV < $\varepsilon$ < 1 eV between Figs. 8(a) and 8(b), though a large positive $\beta$ at $\varepsilon$ < 0.2 eV is not well reproduced in Fig. 8(b) because of the plane-wave approximation. For the second band initially centered at $\varepsilon \sim 2.2$ eV, the value of $\beta$ is slightly negative in Fig. 8(a) and about −0.23 in Fig. 8(b), which is again within the range of the error. At $\varepsilon$ > 3.2 eV where the third band is located, $\beta$ is strongly negative in both Figs. 8(a) and 8(b). Overall the experimental and calculated time-energy 2D maps of $\beta$ are qualitatively consistent with each other. The similarities between them also support our previous finding in Refs. 33 and 37 that ultrafast IC in photo-excited pyrazine predominantly proceeds from $S_2$ to $S_1$, not to other dark $n\pi^*$ states.

**IV. CONCLUSION**

In Ref. 37, we demonstrated that the peak positions and intensities of the VUV PESs predicted for the ionizations of $S_3$ and $S_4$ were totally different from those observed by Horio *et al.*[34], thereby supporting the conventional direct two-state ($S_2 \rightarrow S_1$) picture of ultrafast IC in pyrazine. In this study, to reinforce our previous findings, we have simulated its *time-resolved* VUV PES and PAD (anisotropy parameter) for a more detailed comparison with the experimental ones. To go beyond a single-point calculation under the sudden-ionization approximation adopted in Ref. 37, the results of our quantum nuclear WP dynamics simulation in the 2D space of $Q_{6a}$ and $Q_{10a}$[33] were utilized as in Eq. (6). The time-resolved PES from the S$_1$ and S$_2$ WPs obtained at the MRCISD level reproduced almost all the characteristic features of the measured PES such as a dramatic change from a three-band to two-band structure. We identified ionization channels constituting each spectral band and discovered $1^2B_u \leftarrow$ S$_2$ and $2^2B_g \leftarrow$ S$_1$, which were absent in the experimental assignments[34] and in our earlier study.[37] The computed time-resolved anisotropy parameter is also in good agreement with the measured one. To sum up, using the same 2D ($Q_{6a}$ and $Q_{10a}$) two-state (S$_1$ and S$_2$) model as in Ref. 33, we achieved a much closer match between the experimental and calculated PESs than in Ref. 37.

Sala *et al.* showed a sharp drop in the XMCQDPT2 energy of S$_3$ along the $Q_{8a}$ (a$_g$) vibrational mode.[27,28] This raises the possibility of a CI between S$_2$ and S$_3$ near the FC position along $Q_{8a}$ even in the case where S$_3$ lies above S$_2$ at the FC position. The time-resolved PES simulation including $Q_{8a}$ together with the $Q_4$ (b$_{2g}$) and $Q_5$ (b$_{2g}$) modes, which couple S$_2$ and S$_3$, is expected to completely solve the problem of ultrafast IC in pyrazine, though the present results convince us that our 2D two-state model is adequate in the time range considered here and captures the main relaxation pathway.

The extension of our protocol for calculating a time-resolved PES and PAD to general polyatomic systems (with more than 2D degrees of freedom) may not be straightforward because of the exponential scaling of usual grid-based quantum dynamics methods with the system size. The MCTDH method, which also scales exponentially but with a smaller base, is able to treat a larger number (tens or even hundreds) of degrees of freedom, though the system Hamiltonian with global or fitted potential energy surfaces must still be prepared beforehand.[58] Aiming at on-the-fly full quantum nuclear WP simulations without pre-computed potential energy surfaces, several groups, including us, have recently developed novel approaches that efficiently update the spatial distribution of grid points or basis functions according to the WP propagation.[23,59–61] Incorporating such efficient quantum dynamics methods into the present computational protocol may enable the application of our time-resolved PES simulation to higher-dimensional nonadiabatic problems.

Time-resolved pump-probe photoelectron spectroscopic measurements have recently been conducted with an extreme UV probe pulse to track the entire process of ultrafast radiationless decay ultimately to the ground state in photochemical reactions subject to bond formation or cleavage.[62,63] Simulations of time-resolved PESs incorporated into *ab initio* quantum or semiclassical nonadiabatic molecular dynamics methods are promising to reveal the dynamics of full valence (and even inner-shell) electrons and highly excited vibrations involved in such experiments.

**ACKNOWLEDGMENTS**

M. K. is grateful to Professors T. Horio and T. Suzuki for providing us with their experimental data. F. R. and B. M. acknowledge support from the Fonds National de la Recherche Scientifique (Belgium) F.R.S.-FNRS research grants #T.0132.16 and #J.0012.18 and computational resources grant #2.5020.11 [Consortium des Equipement de Calcul Intensifs (CECI)]. This work was supported in part by JSPS KAKENHI (Grant Number JP18K05022) and by the Japan-Belgium Research Cooperative Program between JSPS and F.R.S.-FNRS (Grant Number JPJSBP120192201).

**DATA AVAILABILITY**

The data that support the findings of this study are available from the corresponding author upon reasonable request.

[1] S. Ullrich, T. Schultz, M. Z. Zgierski, and A. Stolow, J. Am. Chem. Soc. **126**, 2262 (2004).

[2] S. Ullrich, T. Schultz, M. Z. Zgierski, and A. Stolow, Phys. Chem. Chem. Phys. **6**, 2796 (2004).

[3] C. Canuel, M. Mons, F. Piuzzi, B. Tardivel, I. Dimicoli, and M. Elhanine, J. Chem. Phys. **122**, 074316 (2005).

[4] H. Saigusa, J. Photochem. Photobiol. C **7**, 197 (2006).

[5] J. Segarra-Martí, T. Tran, and M. J. Bearpark, Phys. Chem. Chem. Phys. **21**, 14322 (2019).

[6] J. Segarra-Martí, T. Tran, and M. J. Bearpark, ChemPhotoChem **3**, 856 (2019).

[7] J. Kommandeur, W. A. Majewski, W. L. Meerts, and D. W. Pratt, Ann. Rev. Phys. Chem. **38**, 433 (1987).

[8] K. K. Innes, I. G. Ross, and W. R. Moomaw, J. Mol. Spectrosc. **132**, 492 (1988).

[9] T. Suzuki, J. Phys. B: At. Mol. Opt. Phys. **47**, 124001 (2014).

[10] I. Yamazaki, T. Murao, T. Yamanaka, and K. Yoshihara, Faraday Discuss. Chem. Soc. **75**, 395 (1983).

[11] R. Schneider and W. Domcke, Chem. Phys. Lett. **150**, 235 (1988).

[12] L. Seidner, G. Stock, A. L. Sobolewski, and W. Domcke, J. Chem. Phys. **96**, 5298 (1992).

[13] C. Woywod, W. Domcke, A. L. Sobolewski, and H.-J. Werner, J. Chem. Phys. **100**, 1400 (1994).

[14] V. Stert, P. Farmanara, and W. Radloff, J. Chem. Phys. **112**, 4460 (2000).

[15] Y. Suzuki, T. Fuji, T. Horio, and T. Suzuki, J. Chem. Phys. **132**, 174302 (2010).

[16] M. Sukharev and T. Seideman, Phys. Rev. Lett. **93**, 093004 (2004).

[17] P. S. Christopher, M. Shapiro, and P. Brumer, J. Chem. Phys. **125**, 124310 (2006).

[18] P. Puzari, B. Sarker, and S. Adhikari, J. Chem. Phys. **125**, 194316 (2006).

[19] I. Burghardt, K. Giri, and G. A. Worth, J. Chem. Phys. **129**, 174104 (2008).

[20] D. V. Shalashilin, J. Chem. Phys. **132**, 244111 (2010).

[21] M. Sala, M. Saab, B. Lasorne, F. Gatti, and S. Guérin, J. Chem. Phys. **140**, 194309 (2014).

[22] M. Saab, M. Sala, B. Lasorne, F. Gatti, and S. Guérin, J. Chem. Phys. **141**, 134114 (2014).

[23] G. W. Richings and S. Habershon, J. Chem. Phys. **148**, 134116 (2018).

[24] U. Werner, R. Mitrić, T. Suzuki, and V. Bonačić-Koutecký, Chem. Phys. **349**, 319 (2008).

[25] G. Tomasello, A. Humeniuk, and R. Mitrić, J. Phys. Chem. A **118**, 8437 (2014).

[26] C.-K. Lin, Y. Niu, C. Zhu, Z. Shuai, and S. H. Lin, Chem. Asian J. **6**, 2977 (2011).

[27] M. Sala, B. Lasorne, F. Gatti, and S. Guérin, Phys. Chem. Chem. Phys. **16**, 15957 (2014).

[28] M. Sala, S. Guérin, and F. Gatti, Phys. Chem. Chem. Phys. **17**, 29518 (2015).

[29] W. Xie, M. Sapunar, N. Došlić, M. Sala, and W. Domcke, J. Chem. Phys. **150**, 154119 (2019).

[30] K. Sun, W. Xie, L. Chen, W. Domcke, and M. F. Gelin, J. Chem. Phys. **153**, 174111 (2020).

[31] I. C. Walker and M. H. Palmer, Chem. Phys. **153**, 169 (1991).

[32] Y. Okuzawa, M. Fujii, and M. Ito, Chem. Phys. Lett. **171**, 341 (1990).

[33] M. Kanno, Y. Ito, N. Shimakura, S. Koseki, H. Kono, and Y. Fujimura, Phys. Chem. Chem. Phys. **17**, 2012 (2015).

[34] T. Horio, R. Spesyvtsev, K. Nagashima, R. A. Ingle, Y. Suzuki, and T. Suzuki, J. Chem. Phys. **145**, 044306 (2016).

[35] J. Cooper and R. N. Zare, J. Chem. Phys. **48**, 942 (1968).

[36] J. O. Johansson and E. E. B. Campbell, Chem. Soc. Rev. **42**, 5661 (2013).

[37] B. Mignolet, M. Kanno, N. Shimakura, S. Koseki, F. Remacle, H. Kono, and Y. Fujimura, Chem. Phys. **515**, 704 (2018).

[38] H.-J. Werner and P. J. Knowles, J. Chem. Phys. **82**, 5053 (1985).

[39] P. J. Knowles and H.-J. Werner, Chem. Phys. Lett. **115**, 259 (1985).

[40] H.-J. Werner and P. J. Knowles, J. Chem. Phys. **89**, 5803 (1988).

[41] P. J. Knowles and H.-J. Werner, Chem. Phys. Lett. **145**, 514 (1988).

[42] P. J. Knowles and H.-J. Werner, Theor. Chim. Acta **84**, 95 (1992).

[43] I. N. Levine, *Quantum Chemistry*, 6th ed. (Prentice Hall, New Jersey, 2009), pp. 471–635.

[44] H.-J. Werner, P. J. Knowles, G. Knizia, F. R. Manby, and M. Schütz, WIREs: Comput. Mol. Sci. **2**, 242 (2012).

[45] H.-J. Werner, P. J. Knowles, G. Knizia, F. R. Manby, M. Schütz, P. Celani, T. Korona, R. Lindh, A. Mitrushenkov, G. Rauhut, K. R. Shamasundar, T. B. Adler, R. D. Amos, A. Bernhardsson, A. Berning, D. L. Cooper, M. J. O. Deegan, A. J. Dobbyn, F. Eckert, E. Goll, C. Hampel, A. Hesselmann, G. Hetzer, T. Hrenar, G. Jansen, C. Köppl, Y. Liu, A. W. Lloyd, R. A. Mata, A. J. May, S. J. McNicholas, W. Meyer, M. E. Mura, A. Nicklass, D. P. O'Neill, P. Palmieri, D. Peng, K. Pflüger, R. Pitzer, M. Reiher, T. Shiozaki, H. Stoll, A. J. Stone, R. Tarroni, T. Thorsteinsson, and M. Wang, MOLPRO, version 2012.1, Cardiff, UK, 2012.

[46] D. Simah, B. Hartke, and H.-J. Werner, J. Chem. Phys. **111**, 4523 (1999).

[47] C. M. Oana and A. I. Krylov, J. Chem. Phys. **131**, 124114 (2009).

[48] G. M. Seabra, I. G. Kaplan, V. G. Zakrzewski, and J. V. Ortiz, J. Chem. Phys. **121**, 4143 (2004).

[49] B. Mignolet, R. D. Levine, and F. Remacle, Phys. Rev. A **86**, 053429 (2012).

[50] M. Deleuze, B. T. Pickup, and J. Delhalle, Mol. Phys. **83**, 655 (1994).

[51] S. Gozem, A. O. Gunina, T. Ichino, D. L. Osborn, J. F. Stanton, and A. I. Krylov, J. Phys. Chem. Lett. **6**, 4532 (2015).

[52] B. T. Pickup, Chem. Phys. **19**, 193 (1977).

[53] E. Bohl, B. Mignolet, J. O. Johansson, F. Remacle, and E. E. B. Campbell, Phys. Chem. Chem. Phys. **19**, 24090 (2017).

[54] R. N. Zare, *Angular Momentum* (Wiley-Interscience, 1988), pp. 77–79.

[55] T. S. Kuhlman, W. J. Glover, T. Mori, K. B. Møller, and T. J. Martínez, Faraday Discuss. **157**, 193 (2012).

[56] Private communication with Professor T. Horio.

[57] T. Horio, Y. Suzuki, and T. Suzuki, J. Chem. Phys. **145**, 044307 (2016).

[58] M. H. Beck, A. Jäckle, G. A. Worth, and H.-D. Meyer, Phys. Rep. **324**, 1 (2000).

[59] K. G. Komarova, F. Remacle, and R. D. Levine, Chem. Phys. Lett. **699**, 155 (2018).

[60] Y. Arai, K. Suzuki, M. Kanno, and H. Kono, Chem. Phys. Lett. **708**, 170 (2018).

[61] T. Murakami and T. J. Frankcombe, J. Chem. Phys. **150**, 144112 (2019).

[62] S. Pathak, L. M. Ibele, R. Boll, C. Callegari, A. Demidovich, B. Erk, R. Feifel, R. Forbes, M. Di Fraia, L. Giannessi, C. S. Hansen, D. M. P. Holland, R. A. Ingle, R. Mason, O. Plekan, K. C. Prince, A. Rouzée, R. J. Squibb, J. Tross, M. N. R. Ashfold, B. F. E. Curchod, and D. Rolles, Nat. Chem. **12**, 795 (2020).

[63] S. Karashima, A. Humeniuk, R. Uenishi, T. Horio, M. Kanno, T. Ohta, J. Nishitani, R. Mitrić, and T. Suzuki, J. Am. Chem. Soc. (in press).

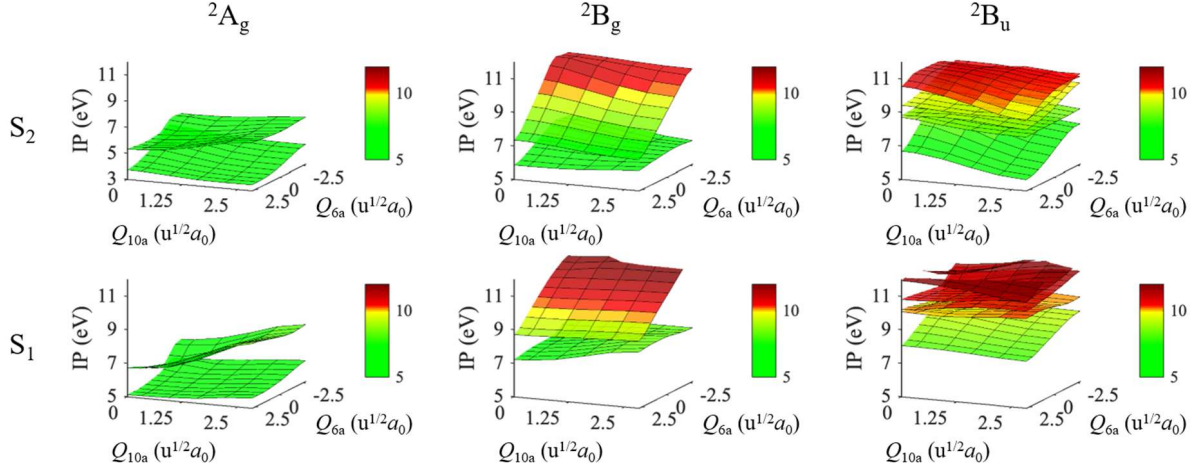

FIG. 1. Vertical IPs for the ionizations from $S_1$ (bottom row) and $S_2$ (top row) to the two $^2A_g$ (left column), two $^2B_g$ (middle column), and five $^2B_u$ (right column) states included in our simulations as a function of the $Q_{6a}$ and $Q_{10a}$ coordinates. They are shifted by 0.53 eV to be closer to experimental data.[33] Note that IPs are plotted from 3 eV in the top left panel but from 5 eV in the other panels.

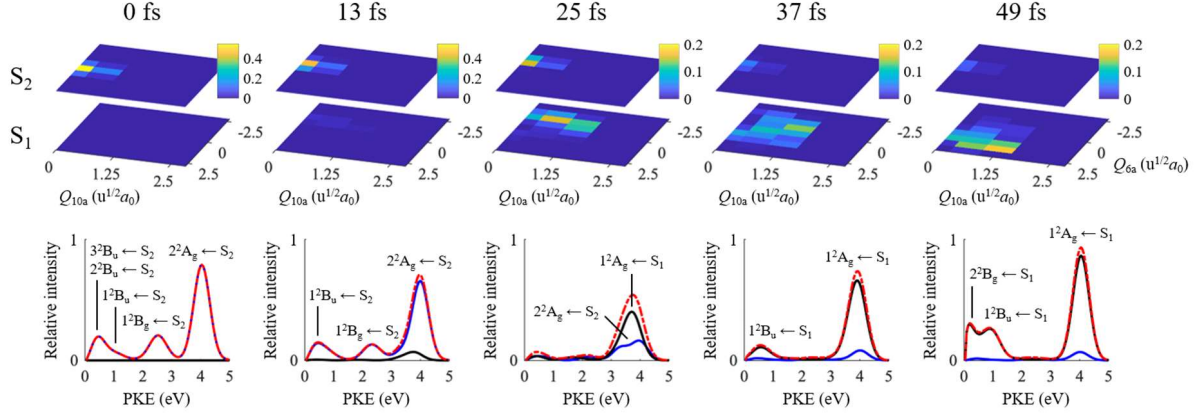

FIG. 2. Spatial localization of nuclear WPs, $C_{Ji}(t)$, of $S_1$ and $S_2$ (top panels)[33] and the sudden-ionization PES from the WPs, $P^{SI}(t, \varepsilon)$, defined in Eq. (7) (bottom panels) at five representative time points ($t$ = 0, 13, 25, 37, and 49 fs). The PESs coming from the ionizations of $S_1$ and $S_2$ are denoted by black and blue solid lines, respectively, and the total PES is represented by a red dashed line. The ionization channel corresponding to each major peak is indicated.

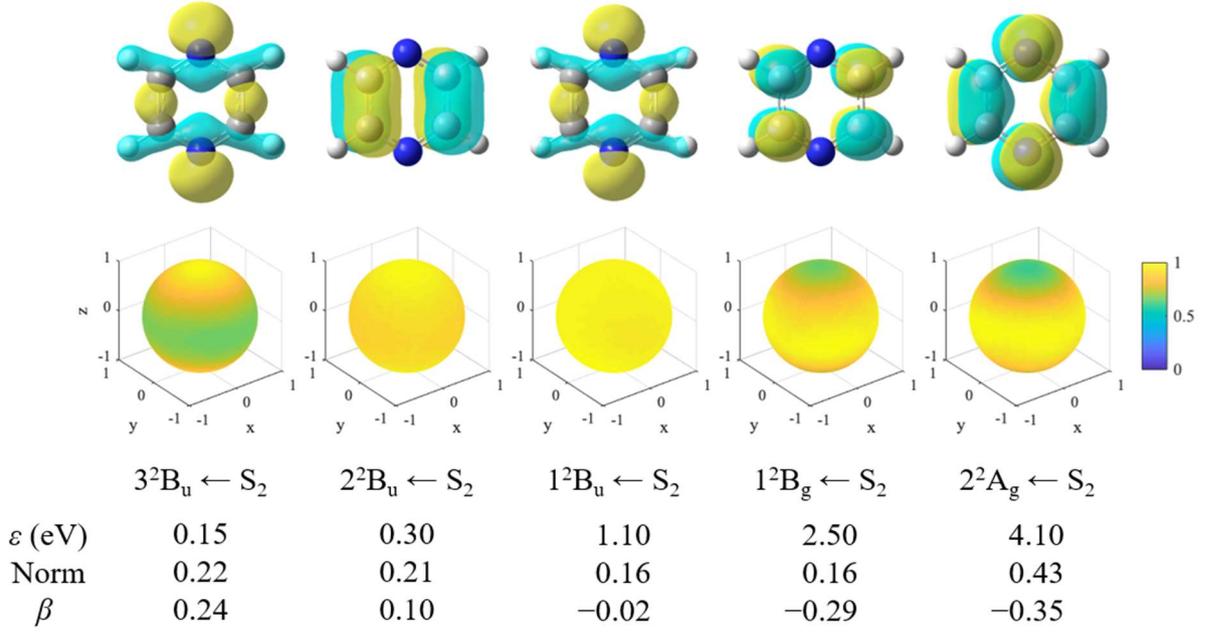

FIG. 3. Dyson orbitals (top panels) and PADs (bottom panels) for the five main ionization channels at $t = 0$. They are shown at the FC position $\mathbf{Q} = 0$ except for the $1^2B_u \leftarrow S_2$ channel shown at $\mathbf{Q} = (0, 0.625\ u^{1/2}a_0)$ because its Dyson orbital norm is almost zero at $\mathbf{Q} = 0$. The white, gray, and blue balls respectively represent hydrogen, carbon, and nitrogen atoms of pyrazine. The values of PKE ($\varepsilon$), Dyson orbital norm, and anisotropy parameter ($\beta$) are presented for each ionization channel. They slightly deviate from the respective values in Table 2 and Fig. 3 in Ref. 37 because of the difference in the number of cationic states included in the SA-CASSCF/MRCISD computation.

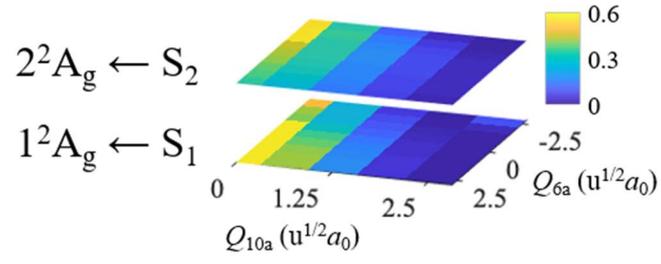

FIG. 4. Norms of the Dyson orbitals for the $2^2A_g \leftarrow S_2$ and $1^2A_g \leftarrow S_1$ channels as a function of the $Q_{6a}$ and $Q_{10a}$ coordinates.

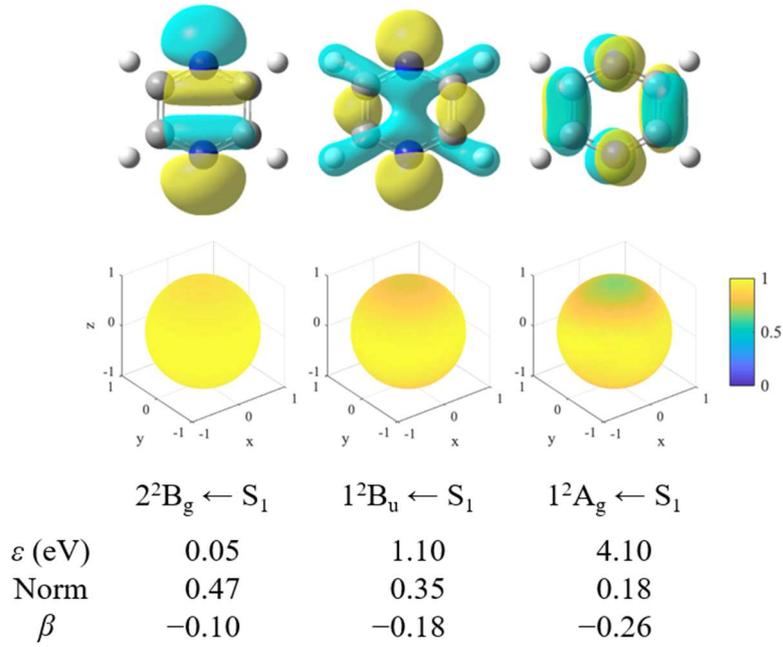

FIG. 5. Dyson orbitals (top panels) and PADs (bottom panels) for the three main ionization channels at $t = 49$ fs. They are shown at $\mathbf{Q} = (1.875\ \mathrm{u}^{1/2}a_0, 1.25\ \mathrm{u}^{1/2}a_0)$, which is the center of the cell with largest probability in $S_1$ (see Fig. 2). The white, gray, and blue balls respectively represent hydrogen, carbon, and nitrogen atoms of pyrazine. The values of PKE ($\varepsilon$), Dyson orbital norm, and anisotropy parameter ($\beta$) are presented for each ionization channel.

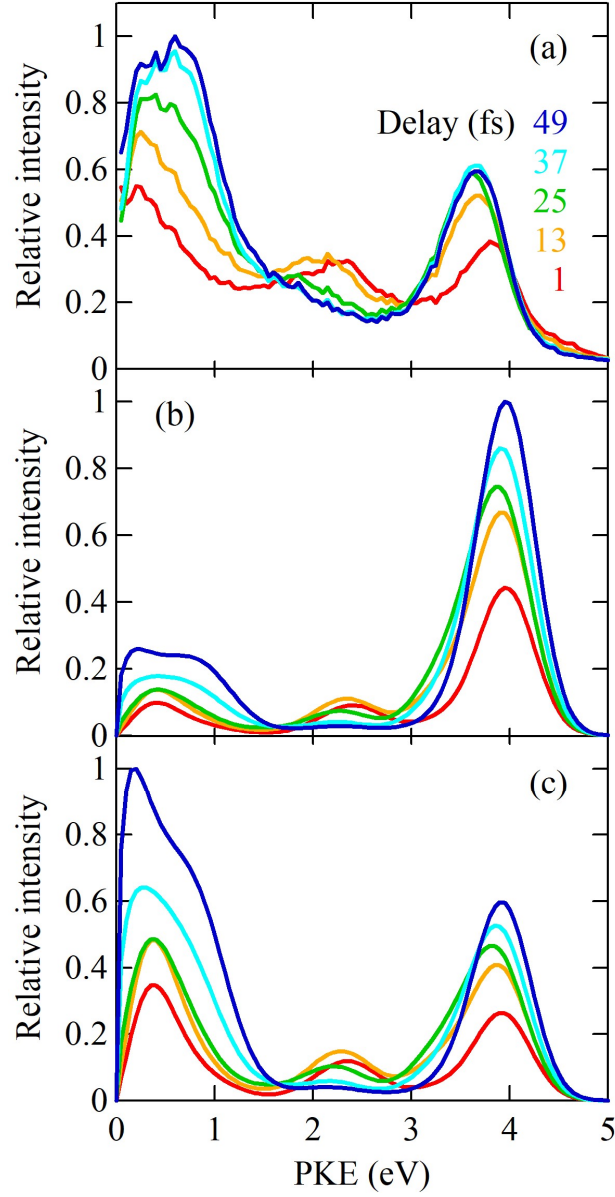

FIG. 6. (a) Experimental time-resolved PES reported in Ref. 34. (b) Calculated time-resolved PES, $P(t, \varepsilon)$, taking into account the temporal broadening induced by pump and probe pulses as in Eq. (8). (c) Product of the PES in panel (b) and $\exp(-\alpha\varepsilon)$ with $\alpha = 0.5$ eV$^{-1}$. The resultant PES is rescaled so that the maximum value is unity.

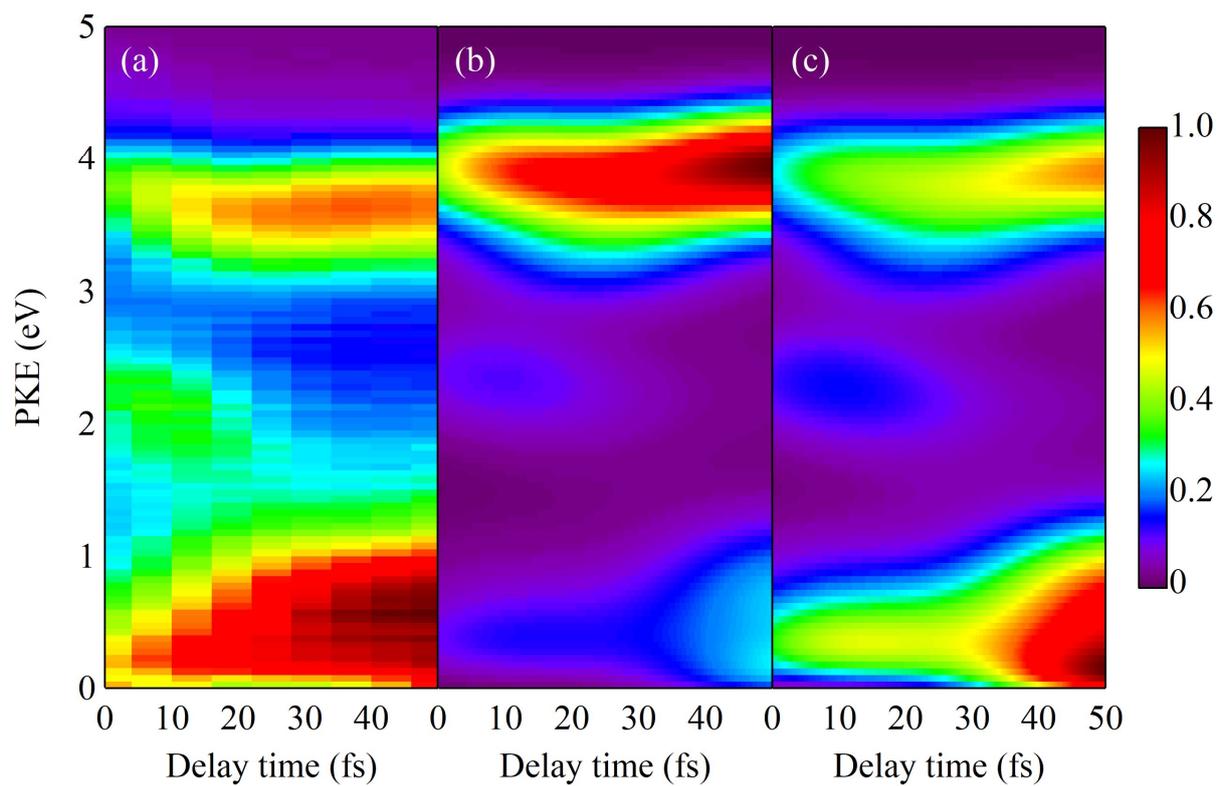

FIG. 7. Time-energy 2D maps of the PESs in Figs. 6(a), 6(b), and 6(c) plotted in panels (a), (b), and (c), respectively.

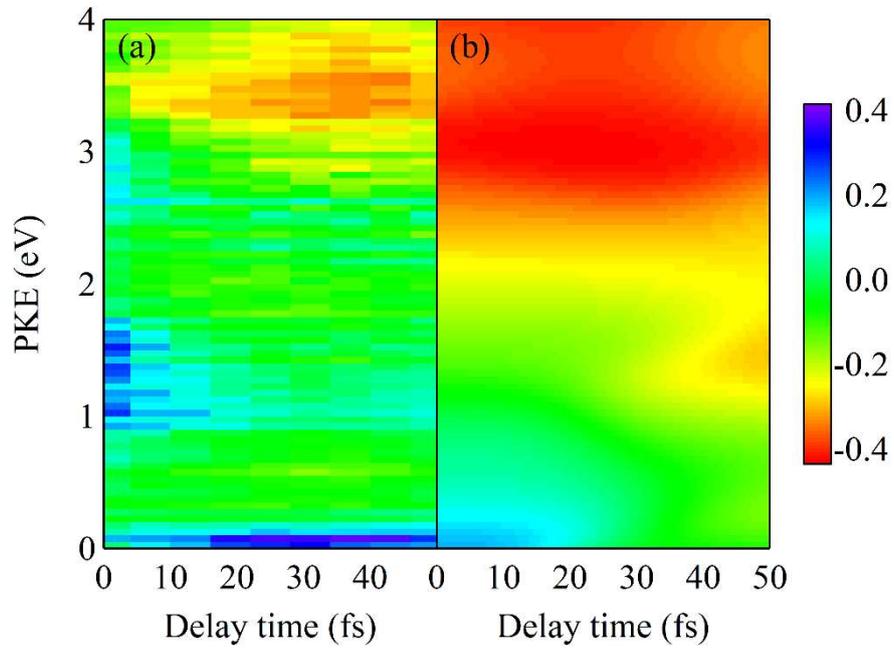

FIG. 8. (a) Experimental[34] and (b) calculated time-energy 2D maps of $\beta$. Effects of pump and probe pulse durations are included in the latter.